\documentclass[journal=jpcl,manuscript=article,layout=traditional]{achemso}
\usepackage{graphicx,dcolumn,bm,xcolor,microtype,multirow,amscd,amsmath,amssymb,amsfonts,physics,mhchem,longtable,xspace,float}
\usepackage{mathpazo,libertine}

\definecolor{darkgreen}{HTML}{009900}
\usepackage[normalem]{ulem}
\newcommand{\titou}[1]{\textcolor{black}{#1}}

\usepackage{hyperref}
\hypersetup{
    colorlinks=true,
    linkcolor=blue,
    filecolor=blue,      
    urlcolor=blue,
    citecolor=blue
}

\newcommand{\tabc}[1]{\multicolumn{1}{c}{#1}}
\newcommand{\SI}{\textcolor{blue}{supporting information}}
\newcommand{\QP}{\textsc{quantum package}}

\newcommand{\ai}[1]{\hat{a}_{#1}}
\newcommand{\aic}[1]{\hat{a}^{\dagger}_{#1}}

\newcommand{\InAA}[1]{#1 \AA}
\newcommand{\kcal}{kcal/mol}

\newcommand{\UEG}{\text{UEG}}
\newcommand{\HF}{\text{HF}}

\newcommand{\PBE}{\text{PBE}}
\newcommand{\FCI}{\text{FCI}}
\newcommand{\CBS}{\text{CBS}}

\newcommand{\CCSDT}{\text{CCSD(T)}}
\newcommand{\lr}{\text{lr}}
\newcommand{\sr}{\text{sr}}

\newcommand{\Ne}{N}

\newcommand{\Nb}{N_{\Bas}}
\newcommand{\Ng}{N_\text{grid}}

\newcommand{\n}[2]{n_{#1}^{#2}}
\newcommand{\Ec}{E_\text{c}}
\newcommand{\E}[2]{E_{#1}^{#2}}
\newcommand{\bE}[2]{\Bar{E}_{#1}^{#2}}

\newcommand{\e}[2]{\varepsilon_{#1}^{#2}}
\newcommand{\be}[2]{\Bar{\varepsilon}_{#1}^{#2}}

\newcommand{\wf}[2]{\Psi_{#1}^{#2}}
\newcommand{\W}[2]{W_{#1}^{#2}}
\newcommand{\w}[2]{w_{#1}^{#2}}

\newcommand{\rsmu}[2]{\mu_{#1}^{#2}}
\newcommand{\V}[2]{V_{#1}^{#2}}
\newcommand{\SO}[2]{\phi_{#1}(\br{#2})}

\newcommand{\Bas}{\mathcal{B}}
\newcommand{\BasFC}{\mathcal{A}}

\newcommand{\Cor}{\mathcal{C}}

\newcommand{\hT}{\Hat{T}}
\newcommand{\hWee}[1]{\Hat{W}_\text{ee}^{#1}}

\newcommand{\f}[2]{f_{#1}^{#2}}
\newcommand{\Gam}[2]{\Gamma_{#1}^{#2}}

\newcommand{\br}[1]{\mathbf{r}_{#1}}
\newcommand{\dbr}[1]{d\br{#1}}

\newcommand{\WFC}[2]{\widetilde{W}_{#1}^{#2}}
\newcommand{\fFC}[2]{\widetilde{f}_{#1}^{#2}}
\newcommand{\rsmuFC}[2]{\widetilde{\mu}_{#1}^{#2}}
\newcommand{\nFC}[2]{\widetilde{n}_{#1}^{#2}}

\newcommand{\LCPQ}{Laboratoire de Chimie et Physique Quantiques (UMR 5626), Universit\'e de Toulouse, CNRS, UPS, France}
\newcommand{\LCT}{Laboratoire de Chimie Th\'eorique, Sorbonne Universit\'e, CNRS, Paris, France}
\newcommand{\ISCD}{Institut des Sciences du Calcul et des Donn\'ees, Sorbonne Universit\'e, Paris, France}

\title{A Density-Based Basis-Set Correction For Wave Function Theory}

\author{Pierre-Fran\c{c}ois Loos}
\email{loos@irsamc.ups-tlse.fr}
\affiliation{\LCPQ}
\author{Bath\'elemy Pradines}
\affiliation{\LCT}
\affiliation{\ISCD}
\author{Anthony Scemama}
\affiliation{\LCPQ}
\author{Julien Toulouse}
\email{toulouse@lct.jussieu.fr}
\affiliation{\LCT}
\author{Emmanuel Giner}
\email{emmanuel.giner@lct.jussieu.fr}
\affiliation{\LCT}

\begin{document}

\newpage
\begin{abstract}
We report a universal density-based basis-set incompleteness correction that can be applied to any wave function method.
The present correction, which appropriately vanishes in the complete basis set (CBS) limit, relies on short-range correlation density functionals (with multi-determinant reference) from range-separated density-functional theory (RS-DFT) to estimate the basis-set incompleteness error.
Contrary to conventional RS-DFT schemes which require an \textit{ad hoc} range-separation \textit{parameter} $\mu$, the key ingredient here is a range-separation \textit{function} $\mu(\bf{r})$ that automatically adapts to the spatial non-homogeneity of the basis-set incompleteness error.
As illustrative examples, we show how this density-based correction allows us to obtain CCSD(T) atomization and correlation energies near the CBS limit for the G2 set of molecules with compact Gaussian basis sets. 
\newpage
\begin{figure}[H]
	\centering
	\includegraphics[width=0.75\linewidth]{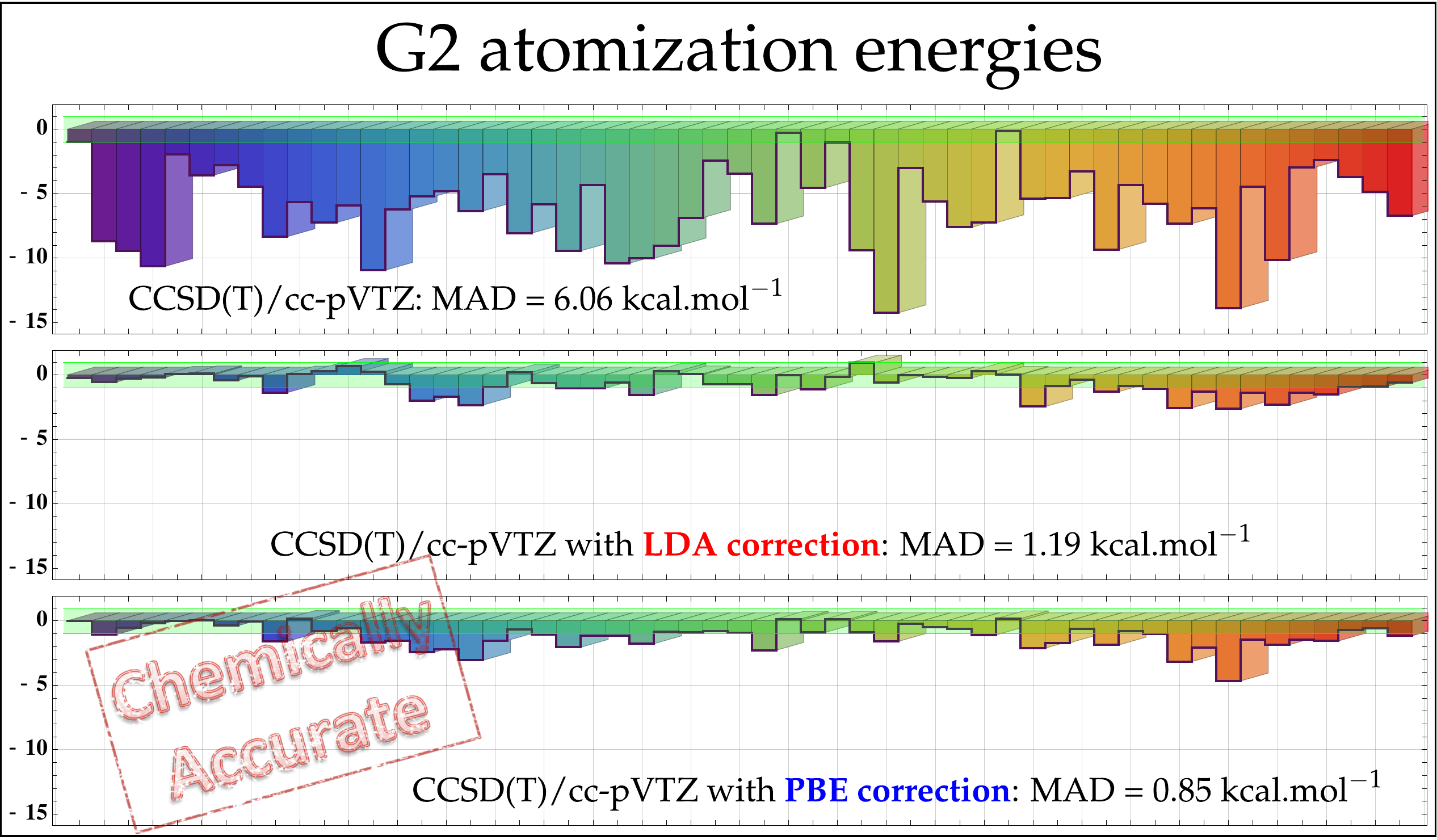}
	\\
	\bf TOC Graphic
\end{figure}
\end{abstract}
\newpage

\maketitle

Contemporary quantum chemistry has developed in two directions --- wave function theory (WFT) \cite{Pop-RMP-99} and density-functional theory (DFT). \cite{Koh-RMP-99}
Although both spring from the same Schr\"odinger equation, each of these philosophies has its own \textit{pros} and \textit{cons}.

WFT is attractive as it exists a well-defined path for systematic improvement as well as powerful tools, such as perturbation theory, to guide the development of new WFT \textit{ans\"atze}.
The coupled cluster (CC) family of methods is a typical example of the WFT philosophy and is well regarded as the gold standard of quantum chemistry for weakly correlated systems. 
By increasing the excitation degree of the CC expansion, one can systematically converge, for a given basis set, to the exact, full configuration interaction (FCI) limit, although the computational cost associated with such improvement is usually high.
One of the most fundamental drawbacks of conventional WFT methods is the slow convergence of energies and properties with respect to the size of the one-electron basis set.
This undesirable feature was put into light by Kutzelnigg more than thirty years ago. \cite{Kut-TCA-85}
To palliate this, following Hylleraas' footsteps, \cite{Hyl-ZP-29} Kutzelnigg proposed to introduce explicitly the interelectronic distance $r_{12} = \abs{\br{1} - \br{2}}$ to properly describe the electronic wave function around the coalescence of two electrons. \cite{Kut-TCA-85, KutKlo-JCP-91, NogKut-JCP-94}
The resulting F12 methods yield a prominent improvement of the energy convergence, and achieve chemical accuracy for small organic molecules with relatively small Gaussian basis sets. \cite{Ten-TCA-12, TenNog-WIREs-12, HatKloKohTew-CR-12, KonBisVal-CR-12, GruHirOhnTen-JCP-17, MaWer-WIREs-18}
For example, at the CCSD(T) level, one can obtain quintuple-$\zeta$ quality correlation energies with a triple-$\zeta$ basis, \cite{TewKloNeiHat-PCCP-07} although computational overheads are introduced by the large auxiliary basis used to resolve three- and four-electron integrals. \cite{BarLoo-JCP-17}
To reduce further the computational cost and/or ease the transferability of the F12 correction, approximated and/or universal schemes have recently emerged. \cite{TorVal-JCP-09, KonVal-JCP-10, KonVal-JCP-11, BooCleAlaTew-JCP-2012, IrmHumGru-arXiv-2019, IrmGru-arXiv-2019}

Present-day DFT calculations are almost exclusively done within the so-called Kohn-Sham (KS) formalism, which corresponds to an exact dressed one-electron theory. \cite{KohSha-PR-65}
The attractiveness of DFT originates from its very favorable accuracy/cost ratio as it often provides reasonably accurate energies and properties at a relatively low computational cost.
Thanks to this, KS-DFT \cite{HohKoh-PR-64, KohSha-PR-65} has become the workhorse of electronic structure calculations for atoms, molecules and solids. \cite{ParYan-BOOK-89}
Although there is no clear way on how to systematically improve density-functional approximations, \cite{Bec-JCP-14} climbing Perdew's ladder of DFT is potentially the most satisfactory way forward. \cite{PerSch-AIPCP-01, PerRuzTaoStaScuCso-JCP-05}
In the context of the present work, one of the interesting feature of density-based methods is their much faster convergence with respect to the size of the basis set. \cite{FraMusLupTou-JCP-15}

Progress toward unifying WFT and DFT are on-going. 
In particular, range-separated DFT (RS-DFT) (see Ref.~\citenum{TouColSav-PRA-04} and references therein) rigorously combines these two approaches via a decomposition of the electron-electron (e-e) interaction into a non-divergent long-range part and a (complementary) short-range part treated with WFT and DFT, respectively. 
As the WFT method is relieved from describing the short-range part of the correlation hole around the e-e coalescence points, the convergence with respect to the one-electron basis set is greatly improved. \cite{FraMusLupTou-JCP-15} 
Therefore, a number of approximate RS-DFT schemes have been developed within single-reference \cite{AngGerSavTou-PRA-05, GolWerSto-PCCP-05, TouGerJanSavAng-PRL-09,JanHenScu-JCP-09, TouZhuSavJanAng-JCP-11, MusReiAngTou-JCP-15} or multi-reference \cite{LeiStoWerSav-CPL-97, FroTouJen-JCP-07, FroCimJen-PRA-10, HedKneKieJenRei-JCP-15, HedTouJen-JCP-18, FerGinTou-JCP-18} WFT approaches.
Very recently, a major step forward has been taken by some of the present authors thanks to the development of a density-based basis-set correction for WFT methods. \cite{GinPraFerAssSavTou-JCP-18}
The present work proposes an extension of this new methodological development alongside the first numerical tests on molecular systems.

The present basis-set correction relies on the RS-DFT formalism to capture the missing part of the short-range correlation effects, a consequence of the incompleteness of the one-electron basis set. 
Here, we only provide the main working equations. 
We refer the interested reader to Ref.~\citenum{GinPraFerAssSavTou-JCP-18} for a more formal derivation.  

Let us assume 
\titou{that we have reasonable approximations of the FCI energy and density of a $\Ne$-electron system in an incomplete basis set $\Bas$, say the CCSD(T) energy $\E{\CCSDT}{\Bas}$ and the Hartree-Fock (HF) density $\n{\HF}{\Bas}$. 
According to Eq.~(15) of Ref.~\citenum{GinPraFerAssSavTou-JCP-18}, the exact ground-state energy $\E{}{}$ may be approximated as}
\begin{equation}
	\label{eq:e0basis}
	\titou{\E{}{} 
	\approx \E{\CCSDT}{\Bas} 
	+ \bE{}{\Bas}[\n{\HF}{\Bas}],}
\end{equation}
where 
\begin{equation}
	\label{eq:E_funcbasis}
	 \bE{}{\Bas}[\n{}{}]
	= \min_{\wf{}{} \to \n{}{}}  \mel*{\wf{}{}}{\hT + \hWee{}}{\wf{}{}} 
	- \min_{\wf{}{\Bas} \to \n{}{}}  \mel*{\wf{}{\Bas}}{\hT + \hWee{}}{\wf{}{\Bas}}
\end{equation}
is the basis-dependent complementary density functional, $\hT$ is the kinetic operator and $\hWee{} = \sum_{i<j} r_{ij}^{-1}$ is the interelectronic repulsion operator.
In Eq.~\eqref{eq:E_funcbasis}, $\wf{}{\Bas}$ and $\wf{}{}$ are two general $\Ne$-electron normalized wave functions belonging to the Hilbert space spanned by $\Bas$ and the complete basis set (CBS), respectively.
Both wave functions yield the same target density $\n{}{}$ (assumed to be representable in $\Bas$). 
Importantly, in the CBS limit (which we refer to as $\Bas \to \CBS$), we have, for any density $\n{}{}$, $\lim_{\Bas \to \CBS} \bE{}{\Bas}[\n{}{}] = 0$. 
This implies that
\begin{equation}
  \label{eq:limitfunc}
        \titou{\lim_{\Bas \to \CBS} \qty( \E{\CCSDT}{\Bas} + \bE{}{\Bas}[\n{\HF}{\Bas}] ) = \E{\CCSDT}{\CBS} \approx \E{}{},}
\end{equation}
where \titou{$\E{\CCSDT}{\CBS}$ is the $\CCSDT$ energy} in the CBS limit.
\titou{Of course, the above holds true for any method that provides a good approximation to the energy and density, not just CCSD(T) and HF.}
In the case where \titou{$\CCSDT$ is replaced by $\FCI$} in Eq.~\eqref{eq:limitfunc}, we have a strict equality as $\E{\FCI}{\CBS} = \E{}{}$.
Provided that the functional $\bE{}{\Bas}[\n{}{}]$ is known exactly, the only sources of error at this stage lie in the approximate nature of the \titou{$\CCSDT$ and $\HF$ methods}, and the lack of self-consistency of the present scheme.

The functional $\bE{}{\Bas}[\n{}{}]$ is obviously \textit{not} universal as it depends on $\Bas$. 
Moreover, as $\bE{}{\Bas}[\n{}{}]$ aims at fixing the incompleteness of $\Bas$, its main role is to correct 
for the lack of cusp (i.e.~discontinuous derivative) in $\wf{}{\Bas}$ at the e-e coalescence points, a universal condition of exact wave functions. 
Because the e-e cusp originates from the divergence of the Coulomb operator at $r_{12} = 0$, a cuspless wave function could equivalently originate from a Hamiltonian with a non-divergent two-electron interaction at coalescence. 
Therefore, as we shall do later on, it feels natural to  approximate $\bE{}{\Bas}[\n{}{}]$ by a short-range density functional which is complementary to a non-divergent long-range interaction.
Contrary to the conventional RS-DFT scheme which requires a range-separation \textit{parameter} $\rsmu{}{}$, here we use a range-separation \textit{function} $\rsmu{}{\Bas}(\br{})$ that automatically adapts to quantify the incompleteness of $\Bas$ in $\mathbb{R}^3$.  

The first step of the present basis-set correction consists in obtaining an effective two-electron interaction $\W{}{\Bas}(\br{1},\br{2})$ ``mimicking'' the Coulomb operator in an incomplete basis $\Bas$. 
In a second step, we shall link $\W{}{\Bas}(\br{1},\br{2})$ to $\rsmu{}{\Bas}(\br{})$.
As a final step, we employ short-range density functionals \cite{TouGorSav-TCA-05} with $\rsmu{}{\Bas}(\br{})$ as range-separation function. 

We define the effective operator as \cite{GinPraFerAssSavTou-JCP-18}
\begin{equation}
	\label{eq:def_weebasis}
	\W{}{\Bas}(\br{1},\br{2})   = 
 	\begin{cases}
       \f{}{\Bas}(\br{1},\br{2})/\n{2}{\Bas}(\br{1},\br{2}), 	& \text{if $\n{2}{\Bas}(\br{1},\br{2}) \ne 0$,}
       \\
       \infty, 												& \text{otherwise,}
    \end{cases}
\end{equation}
where
\begin{equation}
	\label{eq:n2basis}
	\n{2}{\Bas}(\br{1},\br{2})
	= \sum_{pqrs \in \Bas}  \SO{p}{1} \SO{q}{2} \Gam{pq}{rs} \SO{r}{1} \SO{s}{2},
\end{equation}
and $\Gam{pq}{rs} = 2 \mel*{\wf{}{\Bas}}{ \aic{r_\downarrow}\aic{s_\uparrow}\ai{p_\uparrow}\ai{q_\downarrow}}{\wf{}{\Bas}}$ are the opposite-spin pair density associated with $\wf{}{\Bas}$ and its corresponding tensor, respectively, $\SO{p}{}$ is a (real-valued) molecular orbital (MO),
\begin{equation}
	\label{eq:fbasis}
	\f{}{\Bas}(\br{1},\br{2}) 
	= \sum_{pqrstu \in \Bas}  \SO{p}{1} \SO{q}{2} \V{pq}{rs} \Gam{rs}{tu} \SO{t}{1} \SO{u}{2},
\end{equation}
and $\V{pq}{rs}=\langle pq | rs \rangle$ are the usual two-electron Coulomb integrals.
With such a definition, $\W{}{\Bas}(\br{1},\br{2})$ satisfies (see Appendix A of Ref.~\citenum{GinPraFerAssSavTou-JCP-18})
\begin{equation}
	\iint \frac{ \n{2}{\Bas}(\br{1},\br{2})}{r_{12}} \dbr{1} \dbr{2} =
	\iint \W{}{\Bas}(\br{1},\br{2}) \n{2}{\Bas}(\br{1},\br{2}) \dbr{1} \dbr{2},
\end{equation}
which intuitively motivates $\W{}{\Bas}(\br{1},\br{2})$ as a potential candidate for an effective interaction.
Note that the divergence condition of $\W{}{\Bas}(\br{1},\br{2})$ in Eq.~\eqref{eq:def_weebasis} ensures that one-electron systems are free of correction as the present approach must only correct the basis-set incompleteness error originating from the e-e cusp.
As already discussed in Ref.~\citenum{GinPraFerAssSavTou-JCP-18}, $\W{}{\Bas}(\br{1},\br{2})$ is symmetric, \textit{a priori} non translational, nor rotational invariant if $\Bas$ does not have such symmetries. 
Thanks to its definition one can show that (see Appendix B of Ref.~\citenum{GinPraFerAssSavTou-JCP-18}) 
\begin{equation}
	\label{eq:lim_W}
        \lim_{\Bas \to \CBS}\W{}{\Bas}(\br{1},\br{2}) = \frac{1}{r_{12}},
\end{equation}
for any $(\br{1},\br{2})$ such that $\n{2}{\Bas}(\br{1},\br{2}) \ne 0$. 


A key quantity is the value of the effective interaction at coalescence of opposite-spin electrons, $\W{}{\Bas}(\br{},{\br{}})$,
which is necessarily \textit{finite} for an incomplete basis set as long as the on-top pair density $\n{2}{\Bas}(\br{},\br{})$ is non vanishing. 
Because $\W{}{\Bas}(\br{1},\br{2})$ is a non-divergent two-electron interaction, it can be naturally linked to RS-DFT which employs a non-divergent long-range interaction operator.
Although this choice is not unique, we choose here the range-separation function
\begin{equation}
	\label{eq:mu_of_r}
	\rsmu{}{\Bas}(\br{})  = \frac{\sqrt{\pi}}{2} \W{}{\Bas}(\br{},\br{}),
\end{equation}
such that the long-range interaction of RS-DFT, $\w{}{\lr,\mu}(r_{12}) = \erf( \mu r_{12})/r_{12}$, coincides with the effective interaction at coalescence, i.e.~$\w{}{\lr,\rsmu{}{\Bas}(\br{})}(0) = \W{}{\Bas}(\br{},\br{})$ at any $\br{}$.

Once $\rsmu{}{\Bas}(\br{})$ is defined, it can be used within RS-DFT functionals to approximate $\bE{}{\Bas}[\n{}{}]$. 
As in Ref.~\citenum{GinPraFerAssSavTou-JCP-18}, we consider here a specific class of short-range correlation functionals known as correlation energy with multi-determinantal reference (ECMD) whose general definition reads \cite{TouGorSav-TCA-05}
\begin{equation}
  \label{eq:ec_md_mu}
  \bE{\text{c,md}}{\sr}[\n{}{},\rsmu{}{}] 
  = \min_{\wf{}{} \to \n{}{}} \mel*{\Psi}{\hT + \hWee{}}{\wf{}{}}
  - \mel*{\wf{}{\rsmu{}{}}[n]}{\hT + \hWee{}}{\wf{}{\rsmu{}{}}[n]},
\end{equation}
where $\wf{}{\rsmu{}{}}[n]$ is defined by the constrained minimization
\begin{equation}
\label{eq:argmin}
	\wf{}{\rsmu{}{}}[n] = \arg \min_{\wf{}{} \to \n{}{}} \mel*{\wf{}{}}{\hT + \hWee{\lr,\rsmu{}{}}}{\wf{}{}},
\end{equation}
with $\hWee{\lr,\rsmu{}{}} = \sum_{i<j} \w{}{\lr,\rsmu{}{}}(r_{ij})$.
The ECMD functionals admit, for any $\n{}{}$, the following two limits
\begin{align}
	\label{eq:large_mu_ecmd}
	\lim_{\mu \to \infty}  \bE{\text{c,md}}{\sr}[\n{}{},\rsmu{}{}] & = 0,
	&
	\lim_{\mu \to 0}  \bE{\text{c,md}}{\sr}[\n{}{},\rsmu{}{}] & = \Ec[\n{}{}],
\end{align}
where $\Ec[\n{}{}]$ is the usual universal correlation density functional defined in KS-DFT. 
The choice of ECMD in the present scheme is motivated by the analogy between the definition of $\bE{}{\Bas}[\n{}{}]$ [Eq.~\eqref{eq:E_funcbasis}] and the ECMD functional [Eq.~\eqref{eq:ec_md_mu}]. 
Indeed, the two functionals coincide if $\wf{}{\Bas} = \wf{}{\rsmu{}{}}$.
Therefore, we approximate $\bE{}{\Bas}[\n{}{}]$ by ECMD functionals evaluated with the range-separation function $\rsmu{}{\Bas}(\br{})$. 

\titou{Inspired} by the recent functional proposed by some of the authors~\cite{FerGinTou-JCP-18}, we propose here a new Perdew-Burke-Ernzerhof (PBE)-based ECMD functional 
\begin{equation}
	\label{eq:def_pbe_tot}
	\bE{\PBE}{\Bas}[\n{}{},\rsmu{}{\Bas}] =
	\int \n{}{}(\br{}) \be{\text{c,md}}{\sr,\PBE}\qty(\n{}{}(\br{}),s(\br{}),\zeta(\br{}),\rsmu{}{\Bas}(\br{})) \dbr{},
\end{equation}
where \titou{$\zeta = (\n{\uparrow}{} - \n{\downarrow}{})/\n{}{}$ is the spin polarization and} $s=\abs{\nabla \n{}{}}/\n{}{4/3}$ is the reduced density gradient. 
$\be{\text{c,md}}{\sr,\PBE}\qty(\n{}{},s,\zeta,\rsmu{}{})$ interpolates between the usual PBE correlation functional, \cite{PerBurErn-PRL-96} $\e{\text{c}}{\PBE}(\n{}{},s,\zeta)$, at $\rsmu{}{}=0$ and the exact large-$\rsmu{}{}$ behavior, \cite{TouColSav-PRA-04, GorSav-PRA-06, PazMorGorBac-PRB-06} yielding
\begin{subequations}
\begin{gather}
	\label{eq:epsilon_cmdpbe}
	\be{\text{c,md}}{\sr,\PBE}(\n{}{},s,\zeta,\rsmu{}{}) = \frac{\e{\text{c}}{\PBE}(\n{}{},s,\zeta)}{1 + \beta(\n{}{},s,\zeta) \rsmu{}{3} },
	\\
	\label{eq:beta_cmdpbe}
	\beta(\n{}{},s,\zeta) = \frac{3}{2\sqrt{\pi} (1 - \sqrt{2} )} \frac{\e{\text{c}}{\PBE}(\n{}{},s,\zeta)}{\n{2}{\UEG}(\n{}{},\zeta)}.
\end{gather}
\end{subequations}
The difference between the ECMD functional defined in Ref.~\citenum{FerGinTou-JCP-18} and the present expression \eqref{eq:epsilon_cmdpbe}-\eqref{eq:beta_cmdpbe} is that we approximate here the on-top pair density by its \titou{uniform electron gas \cite{LooGil-WIRES-16} (UEG)} version, i.e.~$\n{2}{\Bas}(\br{},\br{})  \approx \n{2}{\UEG}(\n{}{}(\br{}),\zeta(\br{}))$, where $\n{2}{\UEG}(\n{}{},\zeta) \approx \n{}{2} (1-\zeta^2) g_0(n)$ with the parametrization of the UEG on-top pair-distribution function $g_0(n)$ given in Eq.~(46) of Ref.~\citenum{GorSav-PRA-06}. 
This represents a major computational saving without loss of accuracy for weakly correlated systems as we eschew the computation of $\n{2}{\Bas}(\br{},\br{})$.
\titou{The complementary functional $\bE{}{\Bas}[\n{\HF}{\Bas}]$ is approximated by $\bE{\PBE}{\Bas}[\n{\HF}{\Bas},\rsmu{}{\Bas}]$ where $\rsmu{}{\Bas}(\br{})$ is given by Eq.~\eqref{eq:mu_of_r}.}
\titou{The slightly simpler local-density approximation (LDA) version of the ECMD functional is discussed in the {\SI}.}

As most WFT calculations are performed within the frozen-core (FC) approximation, it is important to define an effective interaction within a subset of MOs. 
We then naturally split the basis set as $\Bas = \Cor \bigcup \BasFC$ (where $\Cor$ and $\BasFC$ are the sets of core and active MOs, respectively) and define the FC version of the effective interaction as
 \begin{equation}
 	\label{eq:WFC}
	\WFC{}{\Bas}(\br{1},\br{2})  = 
    \begin{cases}
	\fFC{}{\Bas}(\br{1},\br{2})/\nFC{2}{\Bas}(\br{1},\br{2}),  & \text{if $\nFC{2}{\Bas}(\br{1},\br{2}) \ne 0$},
	\\
	\infty,		&	\text{otherwise,}
    \end{cases}
 \end{equation}
with 
\begin{subequations}
\begin{gather}
	\label{eq:fbasisval}
	\fFC{}{\Bas}(\br{1},\br{2})
	=  \sum_{pq \in \Bas} \sum_{rstu \in \BasFC}  \SO{p}{1} \SO{q}{2} \V{pq}{rs} \Gam{rs}{tu} \SO{t}{1} \SO{u}{2},
	\\
	\nFC{2}{\Bas}(\br{1},\br{2})
	= \sum_{pqrs \in \BasFC}  \SO{p}{1} \SO{q}{2} \Gam{pq}{rs} \SO{r}{1} \SO{s}{2}, 
\end{gather}
\end{subequations}
and the corresponding FC range-separation function $\rsmuFC{}{\Bas}(\br{}) = (\sqrt{\pi}/2) \WFC{}{\Bas}(\br{},\br{})$.
It is noteworthy that, within the present definition, $\WFC{}{\Bas}(\br{1},\br{2})$ still tends to the regular Coulomb interaction as $\Bas \to \CBS$.
\titou{Defining $\nFC{\HF}{\Bas}$ as the FC (i.e.~valence-only) $\HF$ one-electron density in $\Bas$, the FC contribution of the complementary functional is then approximated by $\bE{\PBE}{\Bas}[\nFC{\HF}{\Bas},\rsmuFC{}{\Bas}]$}.

The most computationally intensive task of the present approach is the evaluation of $\W{}{\Bas}(\br{},\br{})$ at each quadrature grid point. 
In the general case (i.e.~$\wf{}{\Bas}$ is a multi-determinant expansion), we compute this embarrassingly parallel step in $\order*{\Ng \Nb^4}$ computational cost with a memory requirement of $\order*{ \Ng \Nb^2}$, where $\Nb$ is the number of basis functions in $\Bas$.
The computational cost can be reduced to $\order*{ \Ng \Ne^2 \Nb^2}$ with no memory footprint when $\wf{}{\Bas}$ is a single Slater determinant.
As shown in Ref.~\citenum{GinPraFerAssSavTou-JCP-18}, this choice for $\wf{}{\Bas}$ already provides, for weakly correlated systems, a quantitative representation of the incompleteness of $\Bas$.
Hence, we will stick to this choice throughout the present study.
In our current implementation, the computational bottleneck is the four-index transformation to get the two-electron integrals in the MO basis which appear in Eqs.~\eqref{eq:n2basis} and \eqref{eq:fbasis}. 
Nevertheless, this step usually has to be performed for most correlated WFT calculations. 

To conclude this section, we point out that, thanks to the definitions \eqref{eq:def_weebasis} and \eqref{eq:mu_of_r} as well as the properties \eqref{eq:lim_W} and \eqref{eq:large_mu_ecmd}, independently of the DFT functional, the present basis-set correction
i) can be applied to any WFT method that provides an energy and a density, 
ii) does not correct one-electron systems, and
iii) vanishes in the CBS limit, hence guaranteeing an unaltered CBS limit for a given WFT method. 


\begin{figure*}
	\includegraphics[width=0.35\linewidth]{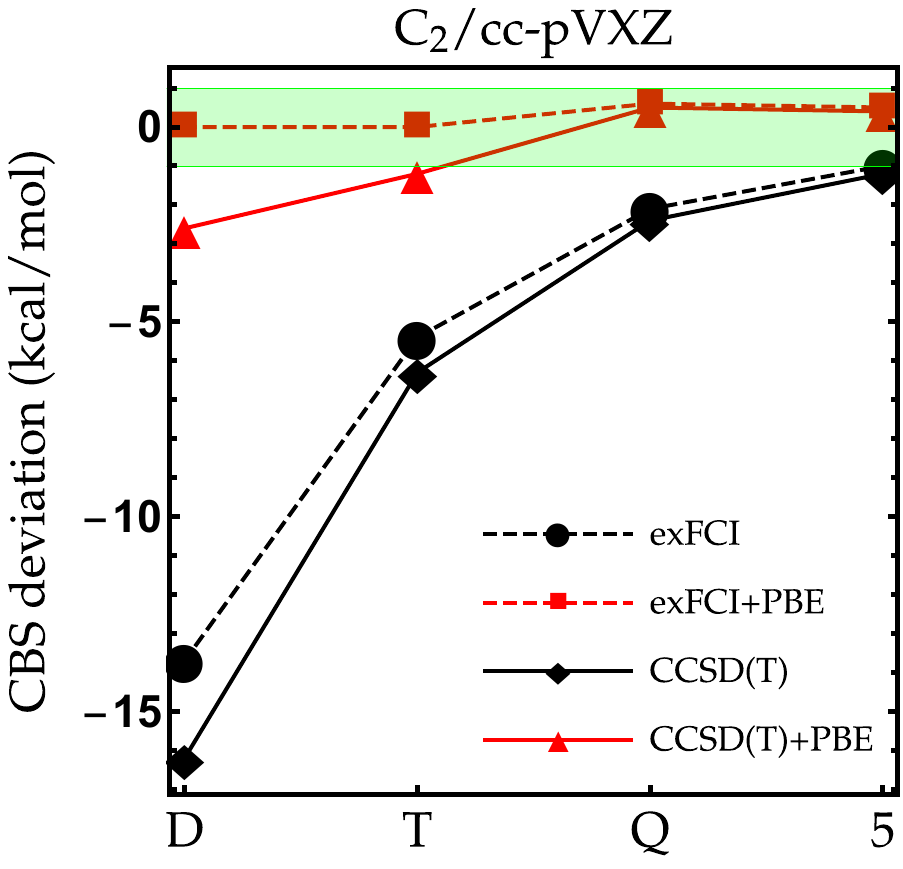}
	\hspace{1cm}
	\includegraphics[width=0.35\linewidth]{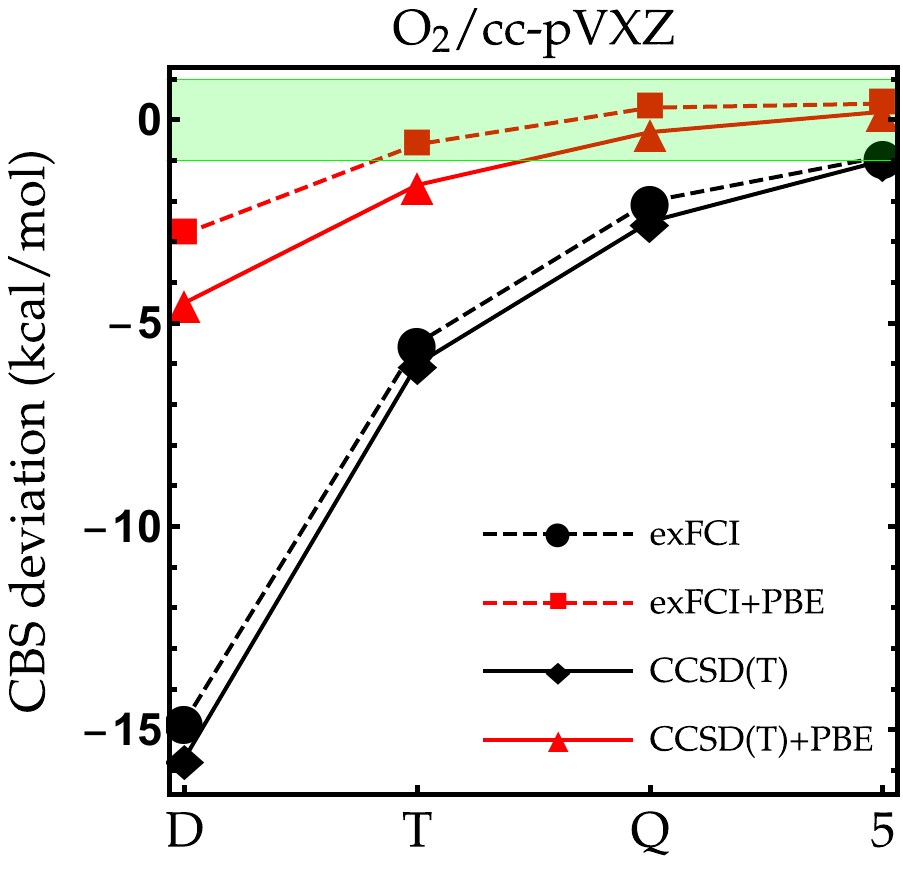}
	\\
	\includegraphics[width=0.35\linewidth]{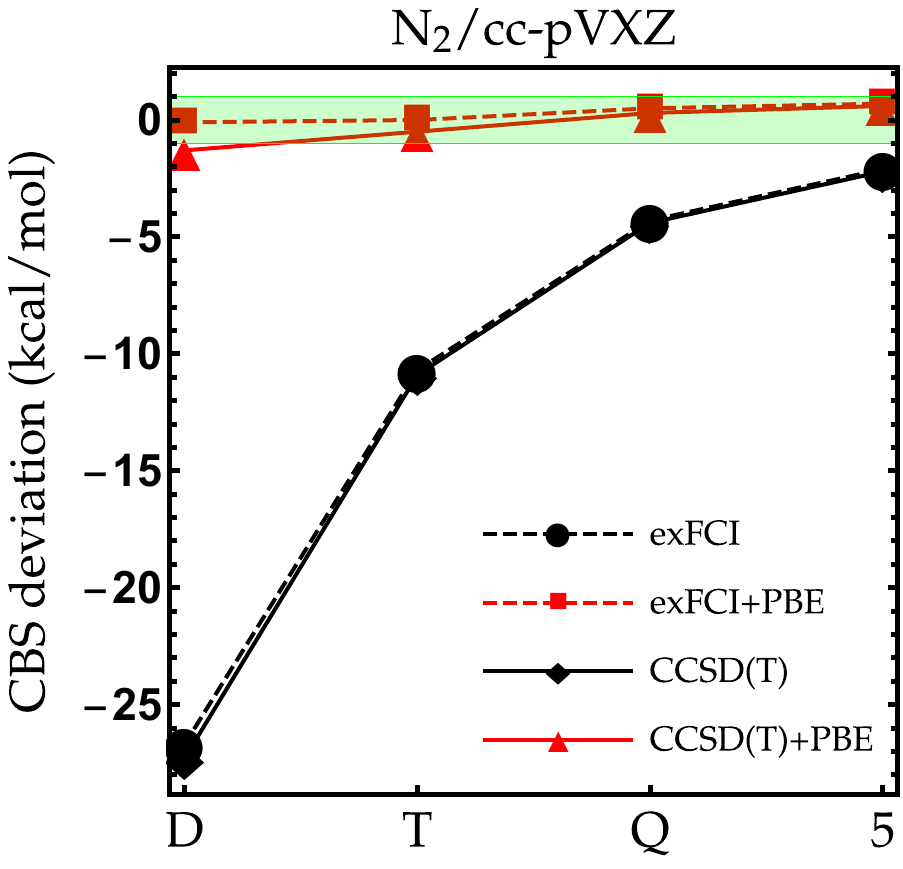}
	\hspace{1cm}
	\includegraphics[width=0.35\linewidth]{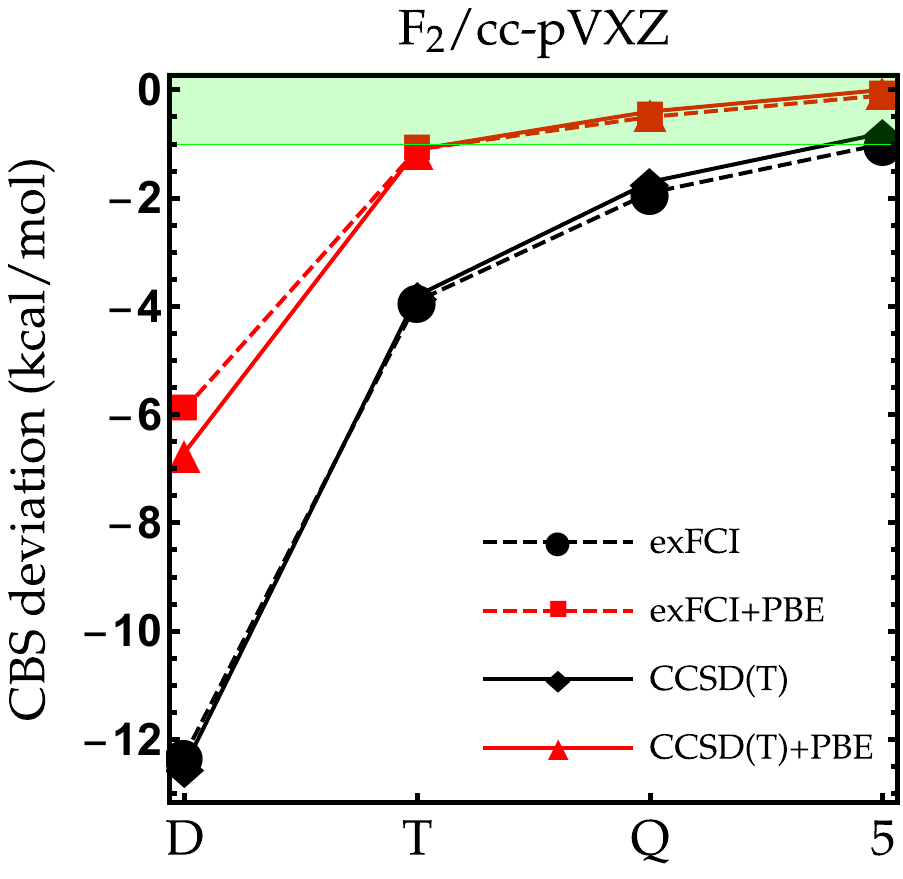}
	\caption{
	Deviation (in \kcal) from CBS atomization energies of \ce{C2} (top left), \ce{O2} (top right), \ce{N2} (bottom left) and \ce{F2} (bottom right) obtained with various methods and basis sets.
	The green region corresponds to chemical accuracy (i.e.~error below 1 {\kcal}).
	See {\SI} for raw data \titou{and the corresponding LDA results}.
	\label{fig:diatomics}}
\end{figure*}

We begin our investigation of the performance of the basis-set correction by computing the atomization energies of \ce{C2}, \ce{N2}, \ce{O2} and \ce{F2} obtained with Dunning's cc-pVXZ basis (X $=$ D, T, Q and 5).
In a second time, we compute the atomization energies of the entire G2 set \cite{CurRagTruPop-JCP-91} composed by 55 molecules with the cc-pVXZ basis set family.
This molecular set has been intensively studied in the last 20 years (see, for example, Refs.~\citenum{FelPetDix-JCP-08, Gro-JCP-09, FelPet-JCP-09, NemTowNee-JCP-10, FelPetHil-JCP-11, HauKlo-JCP-12, PetTouUmr-JCP-12, FelPet-JCP-13, KesSylKohTewMar-JCP-18}) and can be considered as a representative set of small organic and inorganic molecules.
\titou{We employ either CCSD(T) or exFCI to compute the energy of these systems.}
Here, exFCI stands for extrapolated FCI energies computed with the CIPSI algorithm. \cite{HurMalRan-JCP-73, GinSceCaf-CJC-13, GinSceCaf-JCP-15}
We refer the interested reader to Refs.~\citenum{HolUmrSha-JCP-17, SceGarCafLoo-JCTC-18, LooSceBloGarCafJac-JCTC-18, SceBenJacCafLoo-JCP-18, LooBogSceCafJAc-JCTC-19} for more details.
In the case of the CCSD(T) calculations, \titou{we use the restricted open-shell HF (ROHF)} one-electron density to compute the complementary basis-set correction energy. 
In the case of exFCI, the one-electron density is computed from a very large CIPSI expansion containing several million determinants.
CCSD(T) energies are computed with Gaussian09 using standard threshold values, \cite{g09} while RS-DFT and exFCI calculations are performed with {\QP}. \cite{QP2}
For the numerical quadratures, we employ the SG-2 grid. \cite{DasHer-JCC-17}
Apart from the carbon dimer where we have taken the experimental equilibrium bond length (\InAA{1.2425}), all geometries have been extracted from Ref.~\citenum{HauJanScu-JCP-09} and have been obtained at the B3LYP/6-31G(2df,p) level of theory.
Frozen-core calculations are systematically performed and defined as such: a \ce{He} core is frozen from \ce{Li} to \ce{Ne}, while a \ce{Ne} core is frozen from \ce{Na} to \ce{Ar}.
The FC density-based correction is used consistently with the FC approximation in WFT methods.
To estimate the CBS limit of each method, following Ref.~\citenum{HalHelJorKloKocOlsWil-CPL-98}, we perform a two-point X$^{-3}$ extrapolation of the correlation energies using the quadruple- and quintuple-$\zeta$ data that we add up to the HF energies obtained in the largest (i.e.~quintuple-$\zeta$) basis.

As the exFCI atomization energies are converged with a precision of about 0.1 {\kcal}, we can label these as near FCI. 
Hence, they will be our references for \ce{C2}, \ce{N2}, \ce{O2} and \ce{F2}. 
The results for these diatomic molecules are reported in Fig.~\ref{fig:diatomics}. 
The corresponding numerical data \titou{(as well as the corresponding LDA results)} can be found in the {\SI}.
As one can see, the convergence of the exFCI atomization energies is, as expected, slow with respect to the basis set: chemical accuracy (error below 1 {\kcal}) is barely reached for \ce{C2}, \ce{O2} and \ce{F2} even with the cc-pV5Z basis set, and the atomization energies are consistently underestimated.
A similar trend holds for CCSD(T).
Regarding the effect of the basis-set correction, several general observations can be made for both exFCI and CCSD(T). 
First, in a given basis set, the basis-set correction systematically improves the atomization energies.
A small overestimation can occur compared to the CBS value by a few tenths of a {\kcal} (the largest deviation being 0.6 {\kcal} for \ce{N2} at the CCSD(T)+PBE/cc-pV5Z level). 
Nevertheless, the deviation observed for the largest basis set is typically within the CBS extrapolation error, which is highly satisfactory knowing the marginal computational cost of the present correction.
In most cases, the basis-set corrected triple-$\zeta$ atomization energies are on par with the uncorrected quintuple-$\zeta$ ones.

\begin{figure}
	\includegraphics[width=0.75\linewidth]{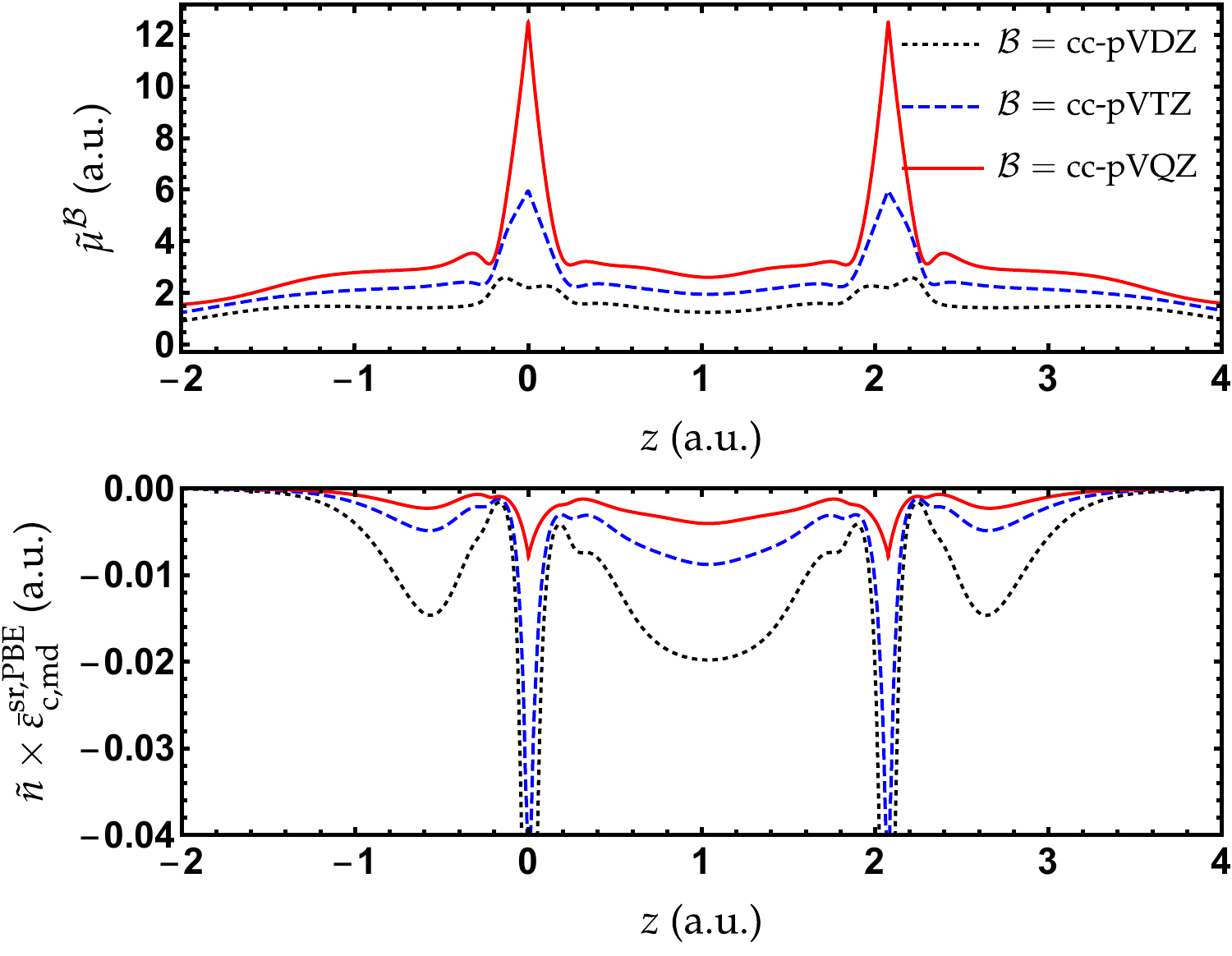}
	\caption{
	$\rsmuFC{}{\Bas}$ (top) and $\nFC{}{} \times \be{\text{c,md}}{\sr,\PBE}$ (bottom) along the molecular axis ($z$) for \ce{N2} for various basis sets.
	The two nitrogen nuclei are located at $z=0$ and $z=2.076$ bohr.
	The calculations have been performed in the FC approximation.
	\label{fig:N2}}
\end{figure}

The fundamental quantity of the present basis-set correction is $\rsmu{}{\Bas}(\br{})$. 
As it grows when one gets closer to the CBS limit, the value of $\rsmu{}{\Bas}(\br{})$ quantifies the quality of a given basis set at a given $\br{}$. 
Another important quantity closely related to $\rsmu{}{\Bas}(\br{})$ is the local energetic correction, $\n{}{}(\br{}) \be{\text{c,md}}{\sr,\PBE}\qty(\n{}{}(\br{}),s(\br{}),\zeta(\br{}),\rsmu{}{\Bas}(\br{}))$, which integrates to the total basis set correction $\bE{\PBE}{\Bas}[\n{}{},\rsmu{}{\Bas}]$ [see Eq.~\eqref{eq:def_pbe_tot}]. 
Such a quantity essentially depends on the local values of both $\rsmu{}{\Bas}(\br{})$ and $\n{}{}(\br{})$.
In order to qualitatively illustrate how the basis-set correction operates, we report, in Figure \ref{fig:N2}, $\rsmuFC{}{\Bas}$ and $\nFC{}{} \times \be{\text{c,md}}{\sr,\PBE}$ along the molecular axis ($z$) of \ce{N2} for $\Bas=\text{cc-pVDZ, cc-pVTZ, cc-pVQZ}$. 
This figure illustrates several general trends: 
i) the value of $\rsmuFC{}{\Bas}(z)$ tends to be much larger than 0.5 bohr$^{-1}$ which is the common value used in RS-DFT, 
ii) $\rsmuFC{}{\Bas}(z)$ is highly non-uniform in space, illustrating the non-homogeneity of basis-set quality in quantum chemistry, 
iii) $\rsmuFC{}{\Bas}(z)$ is significantly larger close to the nuclei, a signature that nucleus-centered basis sets better describe these high-density regions than the bonding regions,
v) the value of the energy correction gets smaller as one improves the basis-set quality, the reduction being spectacular close to the nuclei, and
iv) a large energetic contribution comes from the bonding regions, highlighting the imperfect description of correlation effects in these regions with Gaussian basis sets.

\begin{table}
	\caption{
	Statistical analysis (in \kcal) of the G2 atomization energies depicted in Fig.~\ref{fig:G2_Ec}. 
	Mean absolute deviation (MAD), root-mean-square deviation (RMSD), and maximum deviation (MAX) with respect to the CCSD(T)/CBS reference atomization energies.
	CA corresponds to the number of cases (out of 55) obtained with chemical accuracy.
	See {\SI} for raw data \titou{and the corresponding LDA results}.
	\label{tab:stats}}
		\begin{tabular}{lcccc}
			\hline
			\hline
			Method					&	\tabc{MAD}	&	\tabc{RMSD}		&	\tabc{MAX}	&	\tabc{CA}	\\
			\hline
			CCSD(T)/cc-pVDZ			&	14.29		&	16.21			&	36.95		&	2			\\
			CCSD(T)/cc-pVTZ			&	6.06		&	6.84			&	14.25		&	2			\\
			CCSD(T)/cc-pVQZ			&	2.50		&	2.86			&	6.75		&	9			\\
			CCSD(T)/cc-pV5Z			&	1.28		&	1.46			&	3.46		&	21			\\
			\\
			CCSD(T)+PBE/cc-pVDZ		&	1.96		&	2.59			&	7.33		&	19			\\
			CCSD(T)+PBE/cc-pVTZ		&	0.85		&	1.11			&	2.64		&	36			\\
			CCSD(T)+PBE/cc-pVQZ		&	0.31		&	0.42			&	1.16		&	53			\\
			\hline
			\hline
		\end{tabular}
\end{table}

\begin{figure*}
	\includegraphics[width=\linewidth]{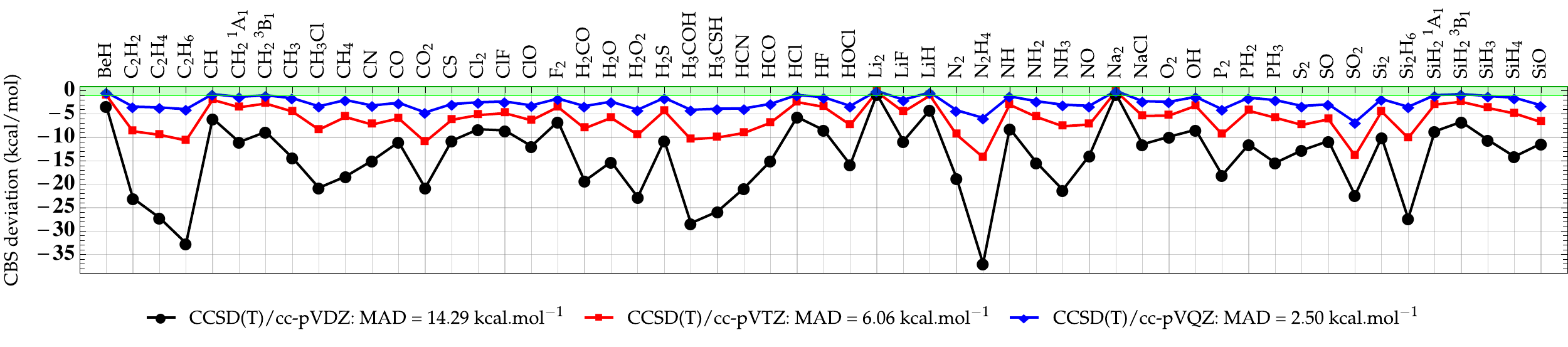}
	\includegraphics[width=\linewidth]{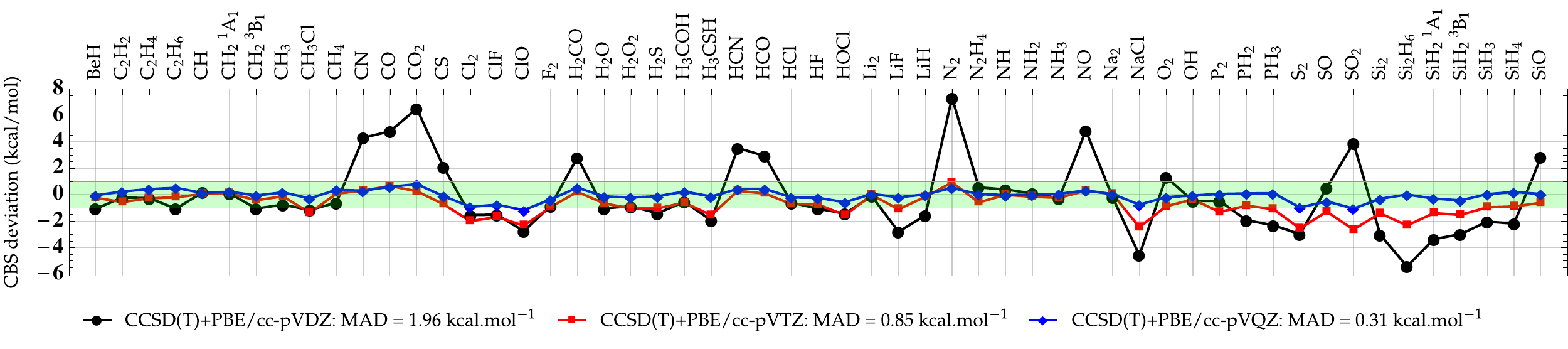}
	\caption{
	Deviation (in \kcal) from the CCSD(T)/CBS atomization energy obtained with \titou{various basis sets for CCSD(T) (top) and CCSD(T)+PBE (bottom).}
	The green region corresponds to chemical accuracy (i.e.~error below 1 {\kcal}).
	\titou{Note the different scales of the vertical axes.}
	See {\SI} for raw data \titou{and the corresponding LDA results}.
	\label{fig:G2_Ec}}
\end{figure*}

As a second set of numerical examples, we compute the error (with respect to the CBS values) of the atomization energies from the G2 test set \titou{with $\CCSDT$} and the cc-pVXZ basis sets.
Here, all atomization energies have been computed with the same near-CBS HF/cc-pV5Z energies; only the correlation energy contribution varies from one method to the other.
Investigating the convergence of correlation energies (or difference of such quantities) is commonly done to appreciate the performance of basis-set corrections aiming at correcting two-electron effects. \cite{Tenno-CPL-04, TewKloNeiHat-PCCP-07, IrmGru-arXiv-2019}
The ``plain'' CCSD(T) atomization energies as well as the corrected \titou{CCSD(T)+PBE} values are depicted in Fig.~\ref{fig:G2_Ec}.
The raw data \titou{(as well as the corresponding LDA results)} can be found in the {\SI}.
A statistical analysis of these data is also provided in Table \ref{tab:stats}, where we report the mean absolute deviation (MAD), root-mean-square deviation (RMSD), and maximum deviation (MAX) with respect to the CCSD(T)/CBS atomization energies.
Note that the MAD of our CCSD(T)/CBS atomization energies is only 0.37 {\kcal} compared to the values extracted from Ref.~\citenum{HauKlo-JCP-12} which corresponds to frozen-core non-relativistic atomization energies obtained at the CCSD(T)(F12)/cc-pVQZ-F12 level of theory corrected for higher-excitation contributions ($E_\text{CCSDT(Q)/cc-pV(D+d)Z} - E_\text{CCSD(T)/cc-pV(D+d)Z})$.
From the double- to the quintuple-$\zeta$ basis, the MAD associated with the CCSD(T) atomization energies goes down slowly from 14.29 to 1.28 {\kcal}.
For a commonly used basis like cc-pVTZ, the MAD of CCSD(T) is still 6.06 {\kcal}.
Applying the basis-set correction drastically reduces the basis-set incompleteness error.
Already at the \titou{CCSD(T)+PBE/cc-pVDZ level}, the MAD is reduced to 1.96 {\kcal}.
With the triple-$\zeta$ basis, the MAD of CCSD(T)+PBE/cc-pVTZ is already below 1 {\kcal} with 36 cases (out of 55) where we achieve chemical accuracy.
\titou{CCSD(T)+PBE/cc-pVQZ returns a MAD of 0.31 kcal/mol} while CCSD(T)/cc-pVQZ still yields a fairly large MAD of 2.50 {\kcal}.

Therefore, similar to F12 methods, \cite{TewKloNeiHat-PCCP-07} we can safely claim that the present basis-set correction provides significant basis-set reduction and recovers quintuple-$\zeta$ quality atomization and correlation energies with triple-$\zeta$ basis sets for a much cheaper computational cost.
Encouraged by these promising results, we are currently pursuing various avenues toward basis-set reduction for strongly correlated systems and electronically excited states.

\section*{Supporting Information Available}
See {\SI} for raw data associated with the atomization energies of the four diatomic molecules and the G2 set \titou{as well as the definition of the LDA ECMD functional (and the corresponding numerical results).}

\begin{acknowledgement}
The authors would like to dedicate the present article to Jean-Paul Malrieu for his 80th birthday.
They would like to thank the \emph{Centre National de la Recherche Scientifique} (CNRS) and the \emph{Institut des Sciences du Calcul et des Donn\'ees} for funding.
This work was performed using HPC resources from GENCI-TGCC (Grant No.~2018-A0040801738) and CALMIP (Toulouse) under allocation 2019-18005. 
\end{acknowledgement}

\bibliography{G2-srDFT,G2-srDFT-control}

\end{document}